\documentstyle[12pt]{article}

\begin{document}

\input{epsf}

\def\beq{\begin{equation}}
\def\eeq{\end{equation}}
\def\bea{\begin{eqnarray}}
\def\eea{\end{eqnarray}}
\def\beas{\begin{eqnarray*}}
\def\eeas{\end{eqnarray*}}
\def\ov{\overline}
\def\ot{\otimes}

\newcommand{\hf}{\mbox{$\frac{1}{2}$}}
\def\sig{\sigma}
\def\De{\Delta}
\def\af{\alpha}
\def\be{\beta}
\def\la{\lambda}
\def\ga{\gamma}
\def\ep{\epsilon}
\def\vep{\varepsilon}
\def\half{\frac{1}{2}}
\def\third{\frac{1}{3}}
\def\fth{\frac{1}{4}}
\def\sth{\frac{1}{6}}
\def\tth{\frac{1}{24}}
\def\tde{\frac{3}{2}}

\def\zb{{\bar z}} 
\def\psib{{\bar \psi}} 
\def\etab{{\bar \eta }}
\def\gab{{\bar \ga}}
\def\vev#1{\langle #1 \rangle}
\def\inv#1{{1 \over #1}}

\def\CA{{\cal A}}       \def\CB{{\cal B}}       \def\CC{{\cal C}}
\def\CD{{\cal D}}       \def\CE{{\cal E}}       \def\CF{{\cal F}}
\def\CG{{\cal G}}       \def\CH{{\cal H}}       \def\CI{{\cal J}}
\def\CJ{{\cal J}}       \def\CK{{\cal K}}       \def\CL{{\cal L}}
\def\CM{{\cal M}}       \def\CN{{\cal N}}       \def\CO{{\cal O}}
\def\CP{{\cal P}}       \def\CQ{{\cal Q}}       \def\CR{{\cal R}}
\def\CS{{\cal S}}       \def\CT{{\cal T}}       \def\CU{{\cal U}}
\def\CV{{\cal V}}       \def\CW{{\cal W}}       \def\CX{{\cal X}}
\def\CY{{\cal Y}}       \def\CZ{{\cal Z}}

\newcommand{\np}{Nucl. Phys.}
\newcommand{\pl}{Phys. Lett.}
\newcommand{\prl}{Phys. Rev. Lett.}
\newcommand{\cmp}{Commun. Math. Phys.}
\newcommand{\jmp}{J. Math. Phys.}
\newcommand{\jpamg}{J. Phys. {\bf A}: Math. Gen.}
\newcommand{\lmp}{Lett. Math. Phys.}
\newcommand{\ptp}{Prog. Theor. Phys.}

\newif\ifbbB\bbBfalse                
\bbBtrue                             

\ifbbB   
 \message{If you do not have msbm (blackboard bold) fonts,}
 \message{change the option at the top of the text file.}
 \font\blackboard=msbm10 
 \font\blackboards=msbm7 \font\blackboardss=msbm5
 \newfam\black \textfont\black=\blackboard
 \scriptfont\black=\blackboards \scriptscriptfont\black=\blackboardss
 \def\Bbb#1{{\fam\black\relax#1}}
\else
 \def\Bbb{\bf}
\fi

\def\bC{{\Bbb C}} 
\def\bZ{{\Bbb Z}}
\def\CN{{\cal N}}

\title{The $su(n)$ Hubbard model}
\author{{\bf Z. Maassarani}\thanks{Work supported by NSERC 
(Canada) and FCAR (Qu\'ebec).} \\
\\
{\small D\'epartement de Physique, Pav. A-Vachon}\\
{\small Universit\'e Laval,  Ste Foy, Qc,  
G1K 7P4 Canada}\thanks{email address: zmaassar@phy.ulaval.ca} \\}
\date{}
\maketitle

\begin{abstract}
The one-dimensional Hubbard model is known to possess
an extended $su(2)$ symmetry and to be integrable. I introduce an
integrable model
with an extended $su(n)$ symmetry. This model contains the usual
$su(2)$ Hubbard model and has a set of features that makes it the 
natural $su(n)$ generalization of the Hubbard model. 
Complete integrability is shown by introducing the $L$-matrix
and showing that the transfer matrix commutes with the hamiltonian.
While the model is integrable in one dimension, it provides a generalization
of the Hubbard hamiltonian in any dimension. 
\end{abstract}
\vspace*{2.5cm}
\noindent
\hspace*{1cm} PACS numbers: 75.10.-b, 75.10.Jm, 75.10.Lp\hfill\\
\hspace*{1cm} Key words: Hubbard model, $su(n)$ spin-chain, integrability

\vspace*{2.5cm}
\noindent
\hspace*{1cm} September 1997\hfill\\
\hspace*{1cm} LAVAL-PHY-24/97\hfill\\
\hspace*{1cm} cond-mat/9709252

\thispagestyle{empty}

\newpage

\setcounter{page}{1}

The two-dimensional Hubbard model \cite{guhu} was first introduced
as a model for the description of the effects of correlation for $d$-electrons
in transition metals.  
The two-dimensional Hubbard model was then shown to be relevant to the study of 
high-$T_c$ superconductivity of cuprate compounds.
Few exact results are known \cite{mont}
and the model is still actively investigated.

In contrast to the two-dimensional model, the one-dimensional
Hubbard model is integrable. However 
ever since the one-dimensional model was recognized as  integrable,  its  peculiar integrable structure still stands alone
outside an integrable hierarchy. 
In this letter I introduce an $n$-state generalized  
model which contains the usual $su(2)$ model. 
This model has been forecast in \cite{mm}.
It is studied in one dimension where it is shown to be integrable. 
Note however that the hamiltonian density is  independent
of the dimension of the lattice
and therefore generalizes the  Hubbard model in any dimension. 

The one-dimensional fermionic $su(2)$ hamiltonian was  first diagonalized
by means of the coordinate Bethe Ansatz \cite{liwu}, a `proof'
of integrability. A Jordan-Wigner transformation shows that the fermionic hamiltonian is  equivalent 
to a bosonic one; 
both forms of the integrable structure were investigated 
in the context of the quantum inverse scattering method \cite{sh12,woa}. 
This algebraic framework unifies integrable one-dimensional systems and
two-dimensional classical statistical mechanics problems \cite{qism}.
The transfer matrix of the latter models  provides  a compact expression
for all the conserved quantities of the quantum models. Showing that 
two such matrices commute directly implies the mutual commutation of the
conserved charges of
the quantum model. Moreover, the method  provides a powerful diagonalization
procedure of all the conserved charges  and is a prerequisite to studying
the model in the thermodynamic limit with the Thermodynamic Bethe Ansatz.

I first introduce the bosonic hamiltonian and the cubic conserved charge.
The two-dimensional `covering' statistical model and  the 
conserved  quantities are then defined. A calculation shows that the 
hamiltonian commutes with  the conserved quantities. Finally some 
comments and possible directions are outlined.

Let $E^{\af\be}$ be the $n\times n$ matrix with a one at row $\af$ 
and column $\be$ and zeros otherwise. 
I define the $su(n)$ Hubbard hamiltonian as:
\bea
H_2 &=&\sum_i h_{ii+1} +\sum_i h^{'}_{ii+1} + U\sum_i h^c_i\\
&=& \sum_i \sum_{\af < n} \left(x E_i^{\af n} E_{i+1}^{n\af} + x^{-1}  
E_i^{n\af} E_{i+1}^{\af n} + (E\rightarrow E^{'})\right) + U
\sum_i (\rho_i +\frac{n-2}{2})  (\rho^{'}_i +\frac{n-2}{2})\nonumber
\eea
where $\rho = \sum_{\af < n} E^{\af\af} -(n-1) E^{nn}$, and 
primed and unprimed quantities correspond to two  commuting   
copies of the $E$ matrices. The complex  free parameter $x$ is a deformation 
inherited from the XX model.  
I am considering
periodic boundary conditions.
The hamiltonians $h$ and $h^{'}$
are just $su(n)$ XX hamiltonians \cite{mm}. These terms correspond
to particle-hopping and the coupling term is an on-site Coulomb-like
interaction. For $|x|=1$ the hamiltonian is hermitian. 
For $n=2$ and $x=1$, and using Pauli matrices, the hamiltonian is just the 
integrable bosonic version of the usual Hubbard hamiltonian \cite{sh12}:
\beq
H_2^{(2)}=\frac{1}{2} \sum_i (\sigma^x_i \sigma^x_{i+1} + \sigma^y_i
\sigma^y_{i+1}) 
+ (\sigma\rightarrow \sigma^{'} ) +  U\sum_i \sigma^z_i\sigma^{'z}_i \nonumber
\eeq
It is possible to write the $n=3$ and $x=1$ hamiltonian 
in terms of  $su(3)$ Gell-Mann matrices:
\beq
H_2^{(3)}= \frac{1}{2} \sum_i \sum_{a\not= 1,2,3,8}
\la^a_i\la^a_{i+1} + (\la\rightarrow \la^{'} ) + 3 U\sum_i (\la^8_i +
\frac{1}{2\sqrt{3}})   (\la^{'8}_i +\frac{1}{2\sqrt{3}}) \nonumber
\eeq

I then constructed a cubic charge which commutes with $H_2$.
Investigation of the $su(3)$ case showed that  two copies of 
the only integrable models which satisfy the Reshetikhin criterion \cite{musc},
can only be coupled in an {\it integrable} way
through $\la^8$. Another argument leading to a single 
coupling along $\rho$ will be the natural generalization of the 
integrability proof.
Let
\beas
h_3 &=& \sum_i \left[ \sum_{\af < n}\left( 
x^2 E_i^{\af n} E_{i+1}^{n n} E_{i+2}^{n\af} -x^{-2} E_i^{n\af} E_{i+1}^{n n} 
E_{i+2}^{\af n}\right) \right.\\
& &\left. +\sum_{\af, \be <n} \left(x^{-2} E_i^{n\af} E_{i+1}^{\af\be} 
E_{i+2}^{\be n} - x^2 E_i^{\af n} E_{i+1}^{\be\af} E_{i+2}^{n\be}
\right) \right]\\
h_3^c &=& n U \sum_i\left[ (x E_i^{\af n} E_{i+1}^{n\af}-x^{-1}
E_i^{n\af} E_{i+1}^{\af n})(\rho^{'}_i + \rho^{'}_{i+1} +n-2) + (E\,\rho^{'}
\rightarrow E^{'}\, \rho ) \right]
\eeas
The conserved cubic hamiltonian is: $H_3 = h_3 + h^{'}_3 + h_3^c$.
A direct calculation shows that $H_2$ and $H_3$ commute. 
The shift of the generator $\rho$ appearing in the coupling terms
is necessary.
The existence of a  cubic conserved quantity is a strong indication 
of integrability; however, as in the $su(2)$ case there is no boost giving
$H_3$ \cite{gp12}.

I now construct the transfer matrix which is the generator of the infinite set 
of conserved quantities.
Consider the $R$-matrix of the $su(n)$ XX model \cite{mm}:
\bea
R(\lambda) &=&\phantom{+} a(\lambda) \; 
[E^{nn}\otimes E^{nn}+\sum_{{\af, \be<n}}
E^{\be \af}\otimes E^{\af\be}] \nonumber\\
& & + b(\lambda)\;\sum_{{\af<n}}(x E^{nn}\otimes E^{\af\af}   + x^{-1} E^{\af\af}\otimes E^{nn})\nonumber\\
& & + c(\lambda)\; \sum_{{\af<n}}(E^{n
\af}\otimes E^{\af n} + E^{\af n}\otimes E^{n\af}) 
\eea
where $a(\la)=\cos(\la)$, $b=\sin(\la)$ and $c(\la)=1$.
The functions $a$, $b$ and $c$ satisfy the `free-fermion' condition:
$a^2 +b^2 = c^2$.
Consider also the matrix 
\beq
I_0 (h)  =\cosh (\frac{h}{2}) + \sinh (\frac{h}{2}) \; C_0 C^{'}_0
\eeq
where $C=\sum_{\af < n} E^{\af\af}-E^{nn}$ and $h$ is so far a free parameter.
The transfer matrix is the trace of a product of $L$-matrices  over 
the auxiliary space 0:
\beq
L_{0i} (\la) = I_0\, R_{0i}\, R^{'}_{0i}\, I_0 \;\;,\;\;\;
\tau (\la)= {\rm Tr}_0 \;\left( L_{0M}...L_{01}\right)
\eeq
where $M$ is the number of sites.
The conserved quantities are given by
\beq
H_{p+1} = \left({d^p \ln\tau (\la)\over d\lambda^p}\right)_{\la=0}
\;\;,\;\; p\geq 0
\eeq
where one drops trivially commuting contributions.
This gives back the quadratic and cubic quantities written above, provided
relation (\ref{rela}) is satisfied. 
The non-existence of a boost, noted earlier, is explained
by the structure of the Lax, $L$,  matrix: $\check{L}^{'\! '}(0) - 
\check{L}^{'2}(0)$ is not proportional to the identity operator 
($\check{L}=P L$ and $P$ is the permutation operator).

The proof that  $H_2$ commutes with  $\tau$ follows the lines 
of the first reference in \cite{sh12}. 
This is one more indication that we are considering the natural multistate
generalization of the  Hubbard model.
The commutator can be written as 
\beq
[ H_2 , \tau (\la) ] = \sum_i {\rm Tr}_0 \left( L_{0M}...[ H_{ii+1},L_{0i+1}
L_{0i} ]... L_{01}\right)\label{com}
\eeq
Let $R_{0ii+1}= [ H_{ii+1},L_{0i+1}L_{0i} ]$, 
where $H_{ii+1}$ is the hamiltonian density.
Let 
\beq
D_{0i}=\frac{c}{a}\left( b\,\frac{\partial}{\partial c}
+c\,\frac{\partial}{\partial b}
\right) R_{0i} R_{0i}^{'}
\eeq
A tedious but straightforward calculation yields  the following 
two equations which are at the root of the integrability proof:
\bea
{[ h_{ii+1},R_{0i+1} ]}R_{0i}+R_{0i+1} {[ h_{ii+1},R_{0i} ]}&=&R_{0i+1}D_{0i}
-D_{0i+1}R_{0i}\\
{[ h_{ii+1},R_{0i+1}]}C_0R_{0i}+R_{0i+1}C_0{[h_{ii+1},R_{0i}]}&=&
-R_{0i+1}C_0D_{0i}+D_{0i+1}C_0R_{0i}
\eea
Using these equations one finds
\beq
R_{0ii+1} = L_{0i+1} Q_{0i}^t -Q_{0i+1} L_{0i}\;\;,\;\;\; 
Q_{0i}= I_0 D_{0i} I_0^{-3} -\frac{n^2 U}{8} [ C_i C^{'}_i , L_{0i} ]
\eeq
Writing the operator $Q$ as a sum of a symmetric piece 
and an antisymmetric piece  one obtains
\bea
&Q_{0i}=A_{0i}-B_{0i}\;\; ,\;\;\; Q_{0i}^t=A_{0i}+B_{0i}&\nonumber\\
&R_{0ii+1} = L_{0i+1} A_{0i} -A_{0i+1} L_{0i} +
L_{0i+1} B_{0i}  + B_{0i+1} L_{0i}&\label{rab} 
\eea
The exact expression of $A$ is not needed while that of $B$ is similar
to that of \cite{sh12}. 
A calculation then shows that for
\beq
\sinh (2h) =\frac{n^2 U}{4}\times\frac{2ab}{c^2} =
\frac{n^2 U}{4} \sin (2\la) \label{rela}
\eeq
one has
\beq
B_{0i}= \left( \frac{c^2+2 b^2}{4 a b}\right) [ L_{0i}, I_0^4 ]
\eeq 
Combining this with eq.~(\ref{rab}), substituting into eq.~(\ref{com}),
and using the cyclic structure of the transfer matrix we find 
$ [ H_2 ,\tau (\la) ]=0$.

We have  shown that all the quantities obtained from the transfer matrix
are conserved thus establishing the integrability of the model.
The conserved quantities will also turn out to commute among themselves.
One just needs to show that two transfer 
matrices at different spectral parameters
commute. This follows trivially if one finds an $R$-matrix intertwining
the monodromy matrices. Then one can start diagonalizing the hamiltonians
by the method of the algebraic Bethe Ansatz.

I conclude with some remarks. The  eigenstate $|n\rangle$  of $E^{nn}$
seems to play a special 
role. However the Hubbard model can be defined with respect to any 
other eigenstate $|\af_0\rangle$, and is unitarily related to the foregoing
model by a unitary transformation built out of the transformation 
exchanging   $|n\rangle$ and $|\af_0\rangle$. 

In one dimension the local symmetry is $(su(n-1)\oplus u(1))\times 
((su(n-1)\oplus u(1))$. This is readily seen by showing that the
operators $\sum_i E_i^{nn}$, and $\sum E_i^{\af\be}$, 
$\af\,,\;\be <n$, commute with the transfer matrix. The primed 
copies are also symmetries. The proof outlined in \cite{mm} holds here.
Finally a transformation of the Jordan-Wigner type (when $n>2$) to fermionic 
variables cannot exist because of dimensional considerations. 
Whether the above local  symmetry  can be extended to  a non-local 
one involving   $E^{\af n}$, $E^{n\af}$, is an open question.

\bigskip\ {\bf Acknowledgement:} I thank P. Mathieu for fruitful 
discussions.

\end{document}